\documentclass[11pt,letterpaper]{article}
\usepackage[T1]{fontenc}
\usepackage[utf8]{inputenc}
\usepackage{lmodern}
\usepackage[margin=1in]{geometry}
\usepackage{amsmath,amssymb,amsthm}
\usepackage{microtype}
\usepackage{xurl}
\usepackage[colorlinks=true,linkcolor=blue,citecolor=blue,urlcolor=blue]{hyperref}
\hypersetup{pdftitle={Pointwise Score Evaluation from Succinct Boolean Descriptions: SAT Equivalence for Strongly Log-Concave Gaussian Mixtures},pdfauthor={Yinzhen Lv}}
\newtheorem{theorem}{Theorem}[section]
\newtheorem{proposition}[theorem]{Proposition}
\newtheorem{corollary}[theorem]{Corollary}
\theoremstyle{definition}
\newtheorem{definition}[theorem]{Definition}
\newtheorem{hypothesis}[theorem]{Hypothesis}
\newcommand{\R}{\mathbb R}
\newcommand{\E}{\mathbb E}
\newcommand{\Cov}{\operatorname{Cov}}
\newcommand{\poly}{\operatorname{poly}}
\newcommand{\PP}{\mathsf P}
\newcommand{\NP}{\mathsf{NP}}
\newcommand{\RP}{\mathsf{RP}}
\newcommand{\norm}[1]{\lVert#1\rVert}
\title{Pointwise Score Evaluation from Succinct Boolean Descriptions:\\ SAT Equivalence for Strongly Log-Concave Gaussian Mixtures}
\author{Yinzhen Lv\\[3pt]
\small Faculty of Engineering and Information Technology (FEIT)\\
\small The University of Melbourne\\
\small\href{mailto:yinzhenlyu@student.unimelb.edu.au}{\texttt{yinzhenlyu@student.unimelb.edu.au}}}
\date{2 November 2026}
\begin{document}
\maketitle
\begin{abstract}
Score-based generative models rely on access to accurate scores, but the cost of computing those scores depends on how the underlying distribution is represented. Can smoothness and strong log-concavity make score evaluation efficient when a distribution is specified by Boolean constraints? I investigate this question by separating geometric encoding from computational extraction. I construct a succinctly specified Gaussian mixture whose score at a fixed query location encodes whether a Boolean formula is satisfiable. The construction has rational component means and uniform strong log-concavity, yet deterministic polynomial-time approximation of the designated score coordinate to constant absolute accuracy is equivalent to polynomial-time solvability of Boolean satisfiability. This equivalence remains valid throughout a neighborhood whose radius is independent of dimension. I further characterize how numerical precision connects bounded witness enumeration to exact counting, and show that independent sample access turns the same target into a dimension-independent mean-estimation problem. An explicit parity construction also separates low-degree approximation from efficient computation. Together, these results show that favorable geometry alone does not resolve the computational cost of extracting scores from compact descriptions. They identify the precise extraction requirement in this formulation while leaving the existence of a general efficient evaluator open.
\end{abstract}

\section{Introduction}
Score-based models use gradients of log densities to describe denoising and sampling procedures. Their analysis distinguishes two tasks: using an accurate score and obtaining that score from the available input. Convergence guarantees address the former under explicit approximation assumptions~\cite{chen}. The latter depends on whether an algorithm receives samples, an explicit mixture, or a compact rule that specifies a distribution.

I study this representation dependence through Boolean satisfiability. A formula compactly specifies a set of satisfying assignments whose size can be exponential in the formula length. I associate these assignments with marked vertices of a hypercube, add a fixed auxiliary atom, and apply unit Gaussian noise. At the origin, all mixture components have the same Gaussian likelihood. The first score coordinate is therefore a rational function of the number of satisfying assignments. Its value separates zero solutions from at least one solution by a constant.

The main characterization holds even under uniform strong log-concavity. I first establish the reduction on hypercube vertices, then compress their coordinates by explicit rational factors. The compressed support lies in a ball of squared radius at most $1/2$. The posterior covariance identity then bounds the Hessian of the negative log density between $I/2$ and $I$ everywhere. An absolute-error guarantee of $1/4$ for the designated score coordinate is still equivalent to $\PP=\NP$ (Theorem~\ref{thm:compressed}). The dimension is $n+1$, the Gaussian noise variance is one, and all curvature constants are independent of $n$.

The access model is part of this statement. With independent samples from the discrete distribution, the target is a bounded mean and is easy to estimate. A formula-to-sampler algorithm would already provide a randomized SAT search procedure. This distinction identifies the computational content of the construction without transferring a pointwise guarantee to an average-error learning problem.

The remaining results refine the characterization. The compressed construction preserves the SAT gap throughout a ball of constant radius. Inverse-polynomial absolute precision remains accessible using polynomially many SAT queries, whereas sufficiently fine precision recovers exact counts. A parity-chain example separates low polynomial degree from efficient computation. Counting-based probability inference and differentiation of compactly specified sums have a classical complexity literature~\cite{roth,darwiche}; my focus is the explicit score embedding and its uniform geometric bounds.

\section{Precise formulation of the motivating hypothesis}
The motivating proposal is to encode Boolean constraints in a continuous space, add noise, and use denoising information to recover a satisfying decision or assignment efficiently. I formalize the fixed-query part of this proposal through two assertions. A geometric encoding specifies where the information is represented. An effective extraction procedure specifies how that information is computed from the formula. These are separate mathematical requirements.

Write $\mathcal F$ for the set of CNF formulas with at least one explicitly represented variable, $n=n(F)$ for the number of these variables, $L=|F|$, and $Z_F$ for the number of satisfying assignments. The compressed density $\widetilde p_F$ and its exact succinct representation are defined in Section~\ref{sec:compressed}. Set $g(F)=[\nabla\log\widetilde p_F(0)]_1$.

\begin{hypothesis}[Geometric encoding assertion $\mathcal G$, proved below]
There is a uniform polynomial-time map from each $F\in\mathcal F$ to a succinct description of a finite rationally supported distribution $\widetilde\mu_F$ on $\R^{n+1}$ such that convolution with $N(0,I_{n+1})$ yields a positive $C^\infty$ density satisfying, for every $y$,
\[
\frac12 I_{n+1}\preceq\nabla^2(-\log\widetilde p_F)(y)\preceq I_{n+1},
\qquad g(F)=\frac{2Z_F-1}{2(2Z_F+1)}.
\]
The description consists of the formula and fixed atom-generation and weighting rules; it does not list the satisfying assignments.
\end{hypothesis}

\begin{hypothesis}[Efficient extraction assertion $\mathcal E$]
For the prescribed compressed family, there exist a deterministic Turing machine $A$ and constants $C>0$ and $k\in\mathbb N$, independent of $F$, such that for every $F\in\mathcal F$, the machine receives only $F$, halts within $C(L+1)^k$ bit operations, and outputs a binary-encoded rational number obeying
\[
\bigl|A(F)-g(F)\bigr|\le\frac14.
\]
All preprocessing and numerical work is included in this bound. The input contains no SAT, counting, score, or target-sampling oracle and no nonuniform advice or unaccounted trained parameters.
\end{hypothesis}

Theorem~\ref{thm:compressed} proves $\mathcal G$ and the equivalence $\mathcal E\Longleftrightarrow\PP=\NP$. It does not establish $\mathcal E$. In particular, efficient construction of a succinct distribution description does not itself provide its normalized probabilities, samples, or score evaluations. For the compressed target, a $1/4$-accurate value is decoded by comparison with $-1/6$; the zero threshold belongs to the uncompressed construction below.

This fixed-query formulation isolates one computational step in the motivating proposal. The paper constructs a fixed noise level, rather than a full reverse-time SDE or probability-flow ODE. A particular reverse-process solver would additionally require a specified initial condition or distribution, dynamics, discretization, termination rule, decoding procedure, and a uniform finite-precision correctness and runtime proof. No such trajectory solver is asserted here. The geometric object proved to exist is a smooth density on Euclidean space; no claim of low intrinsic manifold dimension or unique pathwise noise inversion is needed.

\section{Input model and construction}
Let $F$ be a CNF formula with variables $x_1,\ldots,x_n$, where $n\ge1$, and let $L$ denote its encoding length. Zero-variable formulas can be evaluated directly. Variables are explicitly represented, so $n\le L$ after a conventional encoding choice. Write
\[
S_F=\{x\in\{0,1\}^n:F(x)=1\},\qquad Z_F=|S_F|,\qquad d=n+1.
\]
For $x\in\{0,1\}^n$, define $\sigma(x)_i=2x_i-1$. Introduce the auxiliary point $u=(-1,1,\ldots,1)\in\R^d$ and the marked satisfying-assignment point $v(x)=(1,\sigma(x))$. Define
\begin{equation}\label{eq:measure}
\mu_F=\frac{\delta_u+2\sum_{x\in S_F}\delta_{v(x)}}{1+2Z_F}.
\end{equation}
The auxiliary point ensures that this is a probability measure even when $F$ is unsatisfiable. The first coordinate distinguishes it from every satisfying-assignment atom. All support points have squared norm $d$.

Let $A\sim\mu_F$, let $G\sim N(0,I_d)$ be independent, and put $Y=A+G$. Write the standard Gaussian density as
\[
\varphi_d(z)=(2\pi)^{-d/2}\exp(-\norm{z}^2/2).
\]
The density of $Y$ is
\begin{equation}\label{eq:density}
p_F(y)=\frac{\varphi_d(y-u)+2\sum_{x\in S_F}\varphi_d(y-v(x))}{1+2Z_F}.
\end{equation}
Its score is $s_F(y)=\nabla\log p_F(y)$. The principal target is
\begin{equation}\label{eq:target}
h(F)=[s_F(0)]_1.
\end{equation}
The input to the evaluation problem is $F$, not a list of the atoms in~\eqref{eq:measure}. Equations~\eqref{eq:measure} and~\eqref{eq:density} specify a distribution mathematically; producing their expanded mixtures is not an assumed polynomial-time preprocessing step. Given a candidate assignment, its membership in the support and its atom coordinates are efficiently computable from $F$. Summation over the support is a separate task. I use the ordinary Turing bit model. A rational output is represented by a signed integer numerator and a positive integer denominator, both encoded in binary. Runtime includes preprocessing, arithmetic, and output production. Randomized machines use independent fair random bits.

\begin{definition}[Formula-access score approximation]
For a fixed $\varepsilon>0$, an evaluator receives $F$ and returns a rational number $\widehat h(F)$ satisfying $|\widehat h(F)-h(F)|\le\varepsilon$ for every CNF formula. No sampling, counting, satisfiability, or trained-parameter oracle is included in the input.
\end{definition}

\section{The score identity and SAT equivalence}
\begin{proposition}\label{prop:identity}
The density $p_F$ is positive and $C^\infty$ on $\R^d$. For every $y$,
\begin{equation}\label{eq:identity}
s_F(y)=\E[A\mid Y=y]-y,\qquad \nabla s_F(y)=\Cov(A\mid Y=y)-I_d.
\end{equation}
At the origin,
\begin{equation}\label{eq:origin}
p_F(0)=(2\pi e)^{-d/2},\qquad h(F)=\frac{2Z_F-1}{2Z_F+1}.
\end{equation}
\end{proposition}
\begin{proof}
The mixture in~\eqref{eq:density} is finite and contains a positive-weight Gaussian, establishing positivity and smoothness. Let $w_a$ be its prior atom probabilities. At every $y$, define the posterior version by
\[
\pi_a(y)=\frac{w_a\varphi_d(y-a)}{\sum_b w_b\varphi_d(y-b)},\qquad m(y)=\sum_a\pi_a(y)a.
\]
This supplies an everywhere-defined conditional mean, including individual points of zero probability under the continuous law. Differentiating each Gaussian gives $\nabla\varphi_d(y-a)=(a-y)\varphi_d(y-a)$, and division by $p_F(y)$ yields $s_F(y)=m(y)-y$. Direct differentiation of the posterior weights gives
\[
\partial_j\pi_a(y)=\pi_a(y)(a_j-m_j(y)),\qquad
\partial_jm_i(y)=\sum_a\pi_a(y)a_i a_j-m_i(y)m_j(y).
\]
The latter is the $(i,j)$ entry of the posterior covariance, proving the Jacobian identity.

At $y=0$, every atom contributes the same Gaussian value because $\norm a^2=d$. The normalized density equals that common value, and the posterior equals $\mu_F$. The first-coordinate numerator is $-1+2Z_F$, with denominator $1+2Z_F$.
\end{proof}

\begin{theorem}[Constant-accuracy equivalence]\label{thm:constant}
A deterministic algorithm computes $h(F)$ to absolute error $1/4$ in time polynomial in $L$ for every CNF formula if and only if $\PP=\NP$.
\end{theorem}
\begin{proof}
Suppose such an evaluator exists. If $Z_F=0$, then $h(F)=-1$, and its output is at most $-3/4$. If $Z_F\ge1$, then $h(F)\ge1/3$, and its output is at least $1/12$. Testing whether the output is positive decides SAT in deterministic polynomial time. NP-completeness of SAT gives $\PP=\NP$.

Conversely, assume $\PP=\NP$. A polynomial-time SAT decider finds a satisfying assignment by fixing variables successively and testing the restricted formulas. After finding an assignment $a$, append the blocking clause
\[
B_a=\bigvee_{i:a_i=0}x_i\;\vee\;\bigvee_{i:a_i=1}\neg x_i,
\]
which excludes exactly $a$. Continue until four distinct assignments have been found or the residual formula is unsatisfiable. If the search terminates with fewer than four, the exact count is known, and~\eqref{eq:origin} can be evaluated exactly. If four assignments are found, output $1$. Then
\[
|1-h(F)|=\frac{2}{2Z_F+1}\le\frac29<\frac14.
\]
There are polynomially many SAT calls, and at most four blocking clauses of length $n$ are added. Rational output and all other operations have polynomial bit complexity.
\end{proof}
The theorem is an equivalence about the existence of an evaluator. Under $\PP\ne\NP$, it yields a conditional obstruction to a universal formula-access evaluator. The converse uses SAT decision and bounded witness enumeration; it does not require exact counting for arbitrary $Z_F$.

\section{Geometric regularity and robust query locations}
\begin{proposition}[Global score regularity]\label{prop:regularity}
For every formula, $\norm{\nabla s_F(y)}_{\mathrm{op}}\le d-1=n$ for all $y$. Thus $s_F$ is globally $n$-Lipschitz and, in particular, $d$-Lipschitz.
\end{proposition}
\begin{proof}
For a unit vector $w$,
\[
0\le w^\top\Cov(A\mid Y=y)w\le\E[(w^\top A)^2\mid Y=y]\le d.
\]
The covariance eigenvalues lie in $[0,d]$, so the eigenvalues of the symmetric Jacobian in~\eqref{eq:identity} lie in $[-1,d-1]$. Because $d\ge2$, their absolute values are at most $d-1$. Integrating the Jacobian along a segment proves the Lipschitz bound.
\end{proof}

\begin{corollary}[A neighborhood gap]
If $\norm y\le1/(24d)$, then $[s_F(y)]_1\le-23/24$ for unsatisfiable $F$, whereas $[s_F(y)]_1\ge7/24$ for satisfiable $F$. An approximation with absolute error $1/4$ still distinguishes these cases by its sign.
\end{corollary}
\begin{proof}
Proposition~\ref{prop:regularity} bounds the change from the origin by $1/24$. Apply~\eqref{eq:origin}; after approximation the satisfiable lower bound is $1/24>0$, and the unsatisfiable upper bound is $-17/24<0$.
\end{proof}
The neighborhood result concerns pointwise guarantees in a prescribed region. It does not give a lower bound under $L^2(p_F)$ error, because a small region may have very small probability. Positive density at each point does not supply a dimension-independent probability mass for its neighborhood.

\begin{proposition}[An axis identity]
Let $e_1=(1,0,\ldots,0)$. For every real $r$,
\begin{equation}\label{eq:axis}
[s_F(re_1)]_1+r=\frac{2Z_F e^{2r}-1}{2Z_F e^{2r}+1}.
\end{equation}
\end{proposition}
\begin{proof}
Along this axis, the Gaussian likelihood ratio of any marked atom to the auxiliary atom is $e^{2r}$. The posterior first-coordinate mean therefore has the stated form. Subtracting $r$ gives the score.
\end{proof}
For $r\ge0$, the right side of~\eqref{eq:axis} is $-1$ when $Z_F=0$ and is at least $1/3$ otherwise. The equal-likelihood calculation at the origin is thus one member of a family of explicit query identities.

Smoothness does not imply log-concavity here. Choose a formula with the unique satisfying assignment $x=(0,\ldots,0)$. Its marked atom is $-u$, so the density is a two-component Gaussian mixture with weights $1/3$ and $2/3$. At a location where the posterior component weights are equal,
\[
\nabla^2\log p_F(y)=uu^\top-I_d.
\]
The eigenvalue along $u$ is $d-1$, which is positive for $d>1$ and also shows that the global Lipschitz bound is attained in this family. The next construction removes this non-log-concavity while retaining the complexity equivalence.

\section{Uniform strong log-concavity with the same constant accuracy}\label{sec:compressed}
Define compressed atoms
\begin{equation}\label{eq:atoms}
\widetilde u=\left(-\frac12,\frac{1}{2n},\ldots,\frac{1}{2n}\right),\qquad
\widetilde v(x)=\left(\frac12,\frac{\sigma(x)}{2n}\right).
\end{equation}
Let $\widetilde\mu_F$ have the same weights as~\eqref{eq:measure} at these atoms, and let $\widetilde p_F$ be its convolution with $N(0,I_d)$. This rescales the atoms before adding fresh unit Gaussian noise; the noise covariance is unchanged. Write $\widetilde s_F=\nabla\log\widetilde p_F$, $V_F=-\log\widetilde p_F$, and $g(F)=[\widetilde s_F(0)]_1$. The atoms have equal squared norm
\[
R_n^2=\frac14+\frac{1}{4n}\le\frac12.
\]
Each coordinate in~\eqref{eq:atoms} has $O(\log(n+1))$ bits. The succinct description stores $F$, $n$, and the fixed rules in~\eqref{eq:atoms}, assigning unnormalized weight $1$ to the auxiliary atom and $2F(x)$ to a candidate marked atom. It has polynomial length and can be written in polynomial time. Evaluating an individual candidate weight and its coordinates also takes polynomial time. This is the description-generation requirement in $\mathcal G$; generating a normalized probability table is not part of that requirement.

\begin{theorem}[Strongly log-concave SAT equivalence]\label{thm:compressed}
For every CNF formula, $\widetilde p_F$ is positive and $C^\infty$, and
\begin{equation}\label{eq:curvature}
\frac12 I_d\preceq\nabla^2V_F(y)\preceq I_d\qquad(y\in\R^d).
\end{equation}
Its score is globally $1$-Lipschitz, and the ratio of the stated smoothness and strong-convexity constants is at most $2$. The succinct construction satisfies $\mathcal G$. Moreover, $\mathcal E$ holds if and only if $\PP=\NP$.
\end{theorem}
\begin{proof}
The identities in Proposition~\ref{prop:identity} apply to any finite Gaussian location mixture with unit covariance. If $\widetilde A\sim\widetilde\mu_F$ and $\widetilde Y=\widetilde A+G$, then
\[
\nabla^2V_F(y)=I_d-\Cov(\widetilde A\mid\widetilde Y=y).
\]
Every posterior remains supported in the ball of radius $R_n$. For a unit vector $w$, its variance in direction $w$ is at most $\E[(w^\top\widetilde A)^2\mid\widetilde Y=y]\le R_n^2$. Consequently
\[
(1-R_n^2)I_d\preceq\nabla^2V_F(y)\preceq I_d,
\]
which proves~\eqref{eq:curvature} and the score Lipschitz bound. Equal norms imply that the posterior at the origin is again the prior, so
\begin{equation}\label{eq:compressedtarget}
g(F)=\frac{2Z_F-1}{2(2Z_F+1)}=\frac12 h(F).
\end{equation}
If $F$ is unsatisfiable, an approximation with error $1/4$ is at most $-1/4$. If $F$ is satisfiable, $g(F)\ge1/6$, and such an approximation is at least $-1/12$. Comparing the output with the rational threshold $-1/6$ decides SAT.

Conversely, under $\PP=\NP$, search for at most two distinct satisfying assignments using self-reduction and blocking clauses. If fewer exist, return the exact value of~\eqref{eq:compressedtarget}. If two are found, return $1/2$, whose error is
\[
\frac12-g(F)=\frac{1}{2Z_F+1}\le\frac15<\frac14.
\]
The formula lengths, query count, and rational arithmetic costs are polynomial in $L$.
\end{proof}

\begin{corollary}[A constant-radius neighborhood]
If $\norm y\le1/24$, then every approximation to $[\widetilde s_F(y)]_1$ with absolute error $1/4$ lies below $-1/6$ for unsatisfiable $F$ and above $-1/6$ for satisfiable $F$.
\end{corollary}
\begin{proof}
The first coordinate changes by at most $1/24$ from its value at zero. After adding the approximation error, the unsatisfiable upper bound is $-1/2+1/24+1/4=-5/24$, and the satisfiable lower bound is $1/6-1/24-1/4=-1/8$. Both are separated from $-1/6$ by at least $1/24$.
\end{proof}
The radius and the numerical separation are independent of dimension. This is a uniform pointwise robustness statement; the probability mass of a fixed-radius ball can still decay rapidly in high dimension.

These curvature bounds are compatible with efficient strongly log-concave sampling in oracle models. Such methods evaluate a density, a potential, or a gradient as part of their updates~\cite{dwivedi}. Here an algorithm receives a formula and must implement the designated numerical query from that input. To see the distinction directly, define the unnormalized sum
\[
H_F(y)=e^{y^\top\widetilde u}+2\sum_{x\in S_F}e^{y^\top\widetilde v(x)}.
\]
The equal-norm identity gives
\[
\widetilde p_F(y)=(2\pi)^{-d/2}e^{-\norm y^2/2-R_n^2/2}\frac{H_F(y)}{1+2Z_F},\qquad
\widetilde s_F(y)=-y+\frac{\nabla H_F(y)}{H_F(y)}.
\]
The normalizing constant cancels in the score, while the ratio of the derivative and the sum retains the dependence on satisfying assignments. The input includes a formula; an expanded sum or a tractable circuit representation must be constructed separately. The result concerns the cost of obtaining this query from the compact specification.

\section{Access to samples changes the task}
\begin{proposition}[Sample-access estimation]\label{prop:samples}
Given independent samples $A^{(1)},\ldots,A^{(N)}\sim\mu_F$, the estimator $\widehat h_N=N^{-1}\sum_i A_1^{(i)}$ satisfies
\[
\Pr(|\widehat h_N-h(F)|>\varepsilon)\le2\exp(-N\varepsilon^2/2).
\]
Consequently $N\ge2\varepsilon^{-2}\log(2/\eta)$ suffices for failure probability at most $\eta$.
\end{proposition}
\begin{proof}
At the origin the posterior equals the prior, so $h(F)=\E[A_1]$. Each first coordinate belongs to $[-1,1]$. Hoeffding's inequality gives the result.
\end{proof}
For $\varepsilon=1/4$, the bound is $N\ge32\log(2/\eta)$, independent of dimension. Samples of $Y$ also suffice: $\E[Y_1]=h(F)$, and independence gives a centered moment generating function bounded by $\exp(t^2)$, since the bounded variable $A_1$ and the unit Gaussian each contribute at most $\exp(t^2/2)$. Thus $2\exp(-N\varepsilon^2/4)$ bounds the corresponding sample-mean error probability. Neither estimator requires samples near the origin.

The same observation applies to a standard statistical query interface for $\mu_F$: querying the bounded function $a\mapsto a_1$ at tolerance $\varepsilon$ directly returns an acceptable approximation. This construction therefore yields a formula-access hardness statement, rather than a statistical query lower bound with free access to $\mu_F$.

\begin{proposition}[Formula-to-sampling implication]\label{prop:sampling}
If a uniform randomized algorithm runs in worst-case polynomial time and, on input $F$, returns a discrete candidate whose law has total variation distance at most $1/6$ from $\mu_F$, then $\NP=\RP$. The same implication holds for $\widetilde\mu_F$.
\end{proposition}
\begin{proof}
Generate a candidate, rejecting any malformed encoding. Accept only if it encodes a marked satisfying assignment verified using $F$. For an unsatisfiable formula acceptance is impossible. For a satisfiable formula the target law gives this event probability $2Z_F/(1+2Z_F)\ge2/3$. Total variation error reduces the acceptance probability by at most $1/6$, leaving at least $1/2$. This gives an $\RP$ algorithm for SAT; NP-completeness and $\RP\subseteq\NP$ imply equality. The marked-atom probability is unchanged by compression.
\end{proof}
An exact sampler with expected runtime at most a polynomial $p(L)$ gives the same consequence: truncate it after $12p(L)$ steps and reject on timeout. Markov's inequality bounds the timeout probability by $1/12$, so acceptance on satisfiable inputs is at least $2/3-1/12=7/12$. Exact sampling is stated with expected runtime because a machine using a bounded number of fair bits has dyadic output probabilities, whereas the auxiliary-atom probability $1/(1+2Z_F)$ is non-dyadic when $Z_F>0$.

For the compressed family, $\widetilde A_1\in[-1/2,1/2]$ and $g(F)=\E[\widetilde A_1]$. Hoeffding's inequality gives the dimension-independent bound $2\exp(-2N\varepsilon^2)$ for its sample mean. A statistical query $\widetilde a\mapsto2\widetilde a_1$ at tolerance $2\varepsilon$, followed by division by two, estimates $g(F)$ to tolerance $\varepsilon$. These sample and query statements supply access to the target distribution as an assumption; Proposition~\ref{prop:sampling} addresses the cost of producing that access from $F$.

\section{Precision, bounded enumeration, and exact counting}
\begin{proposition}[A SAT-oracle upper bound]\label{prop:oracle}
For rational $0<\varepsilon\le1/4$, $h(F)$ can be approximated to absolute error $\varepsilon$ using $O(n\lceil1/\varepsilon\rceil)$ SAT queries on formulas of size polynomial in $L+\lceil1/\varepsilon\rceil$. The other bit operations are polynomial in these quantities and the encoding length of $\varepsilon$.
\end{proposition}
\begin{proof}
Repeat the witness search and blocking procedure in Theorem~\ref{thm:constant}, with $K=\lceil1/\varepsilon\rceil$. If all solutions are exhausted before $K$ witnesses are found, evaluate~\eqref{eq:origin} exactly. Otherwise return $1$; its error is at most $2/(2K+1)<\varepsilon$. Each witness uses $O(n)$ SAT queries. The residual formula contains at most $K$ additional clauses of length $n$.
\end{proof}
More precisely, let $\varepsilon(L)$ be a polynomial-time computable rational error schedule such that $0<\varepsilon(L)\le1/4$ and $1/\varepsilon(L)\le P(L)$ for a fixed polynomial $P$. A deterministic polynomial-time evaluator with this error guarantee for every input exists if and only if $\PP=\NP$. The reverse implication follows from Proposition~\ref{prop:oracle}; the forward implication uses the sign gap in Theorem~\ref{thm:constant}. The upper bound is polynomial in $1/\varepsilon$, not necessarily in the number of precision bits.

\begin{proposition}[Fine precision recovers the count]\label{prop:count}
Let $n\ge1$. If $|\widehat h-h(F)|\le\delta_n=2^{-2n-4}$, the exact integer $Z_F$ can be recovered from $\widehat h$ by rational arithmetic and rounding, in time polynomial in its bit length and $n$.
\end{proposition}
\begin{proof}
Write $a=1-h(F)=2/(2Z_F+1)$. Since $Z_F\le2^n$, one has $a>2^{-n-1}$. The assumed error satisfies $\delta_n<a/2$, so $1-\widehat h\ge a/2>0$. For $r(t)=(1+t)/(2(1-t))$,
\[
|r(\widehat h)-r(h(F))|=\frac{|\widehat h-h(F)|}{|1-\widehat h|(1-h(F))}\le\frac{2\delta_n}{a^2}<\frac12.
\]
Because $r(h(F))=Z_F$, rounding recovers the count.
\end{proof}
A uniform evaluator running in $\poly(L,b)$ time for error $2^{-b}$ would consequently compute exact \#SAT counts in polynomial time by taking $b=2n+4$. Here $b$ is a precision parameter encoded in unary; equivalently, the runtime requirement is explicitly polynomial in its value. Conversely, an exact counting algorithm evaluates~\eqref{eq:origin} and rounds it to any requested precision in polynomial time in $L+b$. Thus the precision-bit formulation is tied to exact counting, whereas the constant-error formulation in Theorem~\ref{thm:constant} is tied to SAT decision. Exponentially small absolute error requires a linear number of precision bits in this argument. Since $h(F)=2g(F)$, the compressed construction has the same conclusions after rescaling the requested tolerance by two; error $2^{-2n-5}$ for $g(F)$ suffices for exact count recovery.

The spacing between neighboring count values makes this distinction explicit:
\[
h(Z+1)-h(Z)=\frac{4}{(2Z+1)(2Z+3)}.
\]
When $Z$ is exponential in $n$, inverse-polynomial absolute error cannot distinguish all adjacent counts through direct inversion. This spacing observation concerns information in one approximate value; Proposition~\ref{prop:count} supplies the positive recovery guarantee at finer precision.

\section{An exact low-degree separation on a tractable family}
The low-degree method studies specified polynomial statistics of the input, and its interpretation depends on the model and distribution~\cite{kunisky,holmgren}. The following family gives a direct regression calculation for my score target.

For $k\ge1$ and $b\in\{0,1\}^k$, introduce variables $x_0,\ldots,x_k$ and require
\[
x_0=0,\qquad x_k=0,\qquad x_i=x_{i-1}\oplus b_i\quad(1\le i\le k).
\]
Encode an equality by $(\neg x_{i-1}\vee x_i)\wedge(x_{i-1}\vee\neg x_i)$ and an inequality by $(x_{i-1}\vee x_i)\wedge(\neg x_{i-1}\vee\neg x_i)$. Add unit clauses fixing the endpoints, and call the resulting CNF $F_b$. The clauses remain in a fixed order with variable indices independent of $b$.

The initial value determines every subsequent variable uniquely. Hence $Z_{F_b}=1$ if $\bigoplus_i b_i=0$, and $Z_{F_b}=0$ otherwise. Writing $\chi(b)=(-1)^{\sum_i b_i}$, I obtain
\begin{equation}\label{eq:parity}
h(F_b)=-\frac13+\frac23\chi(b).
\end{equation}

\begin{theorem}[Low-degree regression error]\label{thm:degree}
If $b$ is uniform on $\{0,1\}^k$, every real polynomial $q(b)$ of total degree less than $k$ satisfies
\[
\E_b[(q(b)-h(F_b))^2]\ge\frac49.
\]
The constant polynomial $q=-1/3$ attains equality.
\end{theorem}
\begin{proof}
Every monomial of degree less than $k$ omits at least one input coordinate. Averaging its product with $\chi(b)$ over that coordinate gives zero. This remains true for nonmultilinear monomials because they omit a coordinate whenever their total degree is less than $k$. Thus $\E[q\chi]=0$ and $\E[\chi]=0$. Substituting~\eqref{eq:parity} gives
\[
\E[(q-h)^2]=\E[(q+1/3)^2]+\frac49\E[\chi^2]\ge\frac49.
\]
The equality claim follows immediately.
\end{proof}
The theorem excludes uniform absolute error $1/4$ for this polynomial class, because that guarantee would imply squared error at most $1/16$. For the compressed target, $g(F_b)=-1/6+\chi(b)/3$ and the identical calculation gives minimum squared error $1/9$ at every degree below $k$. Hence uniform error $1/4$ is also excluded in the strongly log-concave family. In the fixed-slot CNF encoding, each varying literal-sign coordinate is a constant, a bit $b_i$, or its complement. Substitution preserves the degree bound, so the result extends to polynomials of those raw sign features.

Nevertheless, computing $\bigoplus_i b_i$ takes $O(k)$ Boolean operations. The degree-$k$ expression $\prod_i(1-2b_i)$ also has a linear-size arithmetic circuit. This example separates polynomial degree from circuit size and runtime. The lower bound concerns direct polynomial regression on specified input features; it does not apply to unrestricted preprocessing, general iterative solvers, or arbitrary neural architectures. In particular, if preprocessing computes parity first, a constant function of the diffusion state can return the correct target.

\section{Relation to score estimation and diffusion theory}
Roth~\cite{roth} relates probabilistic reasoning over propositional descriptions to model counting. Darwiche~\cite{darwiche} represents probabilistic inference through multivariate polynomials and their derivatives, with complexity depending on compilation and evaluation. My reduction uses the same general distinction between a compact logical specification and access to its associated sums. The concrete conclusion is a fixed-point, single-coordinate approximation equivalence that persists under the uniform curvature bounds in~\eqref{eq:curvature}.

Chen et al.~\cite{chen} prove sampling guarantees from accurate score estimates under their stated assumptions. My construction addresses the computation of one score value from a Boolean description. A reduction that assumes accurate scores and an algorithm that constructs those scores are distinct computational objects.

Song~\cite{song} establishes cryptographic hardness of $L^2$-accurate score estimation even when polynomial sample complexity is available. That work concerns a learning-access model and lattice-based hardness assumptions. The present reduction uses worst-case formula access and relates a fixed pointwise approximation guarantee to SAT. These statements address different inputs and error criteria. The sample estimator in Proposition~\ref{prop:samples} explicitly shows why the current construction alone does not establish sample-access score-learning hardness.

Chewi et al.~\cite{chewi} connect score estimation with parameter and PAC density estimation, obtaining computational learning lower bounds that include Gaussian mixtures. Their score-estimation framework uses distributional error along a noising process. My target is a designated scalar value from a known formula specification. Gupta et al.~\cite{gupta} study a further access distinction: posterior sampling can be computationally hard under one-way-function assumptions even when unconditional sampling is efficient. Their posterior task includes observations and a measurement model; my reduction uses a direct formula-to-value interface.

Strongly log-concave sampling guarantees such as those of Dwivedi et al.~\cite{dwivedi} count iterations that use density or gradient queries. Theorem~\ref{thm:compressed} supplies constant curvature bounds while retaining hardness of implementing one such gradient query from the formula. It therefore concerns a different computational resource from the number of iterations once those queries are available.

Kunisky, Wein, and Bandeira~\cite{kunisky} develop the low-degree likelihood-ratio framework for hypothesis testing and distinguish its rigorous results from conjectural computational interpretations. Holmgren and Wein~\cite{holmgren} give counterexamples to a formulation of a low-degree conjecture and analyze the role of its assumptions. Theorem~\ref{thm:degree} is instead a self-contained regression statement under a specified uniform input distribution. It supplies an exact restricted-model result without identifying low degree with all polynomial-time computation.

The construction gives a representation-sensitive characterization: the formula is the computational input, while the exponentially large mixture is a mathematical description. The regularity bounds concern the resulting continuous field, and the complexity equivalence concerns extracting a value from its compact Boolean specification.

\section{Conclusion}
A marked Boolean distribution followed by unit Gaussian smoothing establishes the geometric encoding assertion $\mathcal G$. Compressing the atoms makes the negative log density globally $1/2$-strongly convex and $1$-smooth while preserving a constant-gap satisfiability signal. I prove that the uniform extraction assertion $\mathcal E$ is equivalent to $\PP=\NP$, with robustness on a ball of constant radius. Sample access makes the target a simple mean, and the parity family separates low-degree approximation from runtime. The result identifies an exact computational requirement for the proposed geometric route: the encoding and the readout reduction are proved, while the existence of a general efficient evaluator remains open.

\end{document}